\documentclass[aps,10pt,prb,twocolumn,showpacs,amsmath,amssymb,superscriptaddress]{revtex4}

\usepackage{epstopdf}
\usepackage{commath}
\usepackage{dsfont}

\usepackage{graphicx}
\usepackage{hyperref}
\usepackage{amssymb}
\usepackage{placeins}
\usepackage{amsmath}
\usepackage{float}

\usepackage{enumitem}
\usepackage{booktabs}
\usepackage{cleveref}

\usepackage{color,soul}

\usepackage[caption=false]{subfig}

\usepackage[final]{pdfpages}

\usepackage{array}

\begin{document}

\title{Switchable valley filter based on a graphene $p$-$n$ junction in a magnetic field}

\author{T. Sekera\footnote{Correspondence address: tibor.sekera@unibas.ch}}
\affiliation{Department of Physics, University of Basel,
Klingelbergstrasse 82, CH-4056 Basel, Switzerland}
\author{C. Bruder}
\affiliation{Department of Physics, University of Basel,
Klingelbergstrasse 82, CH-4056 Basel, Switzerland}
\author{E.J. Mele}
\affiliation{Department of Physics and Astronomy, University of Pennsylvania, Philadelphia PA 19104, USA}
\author{R.P. Tiwari}
\affiliation{Department of Physics, University of Basel,
Klingelbergstrasse 82, CH-4056 Basel, Switzerland}
\affiliation{Department of Physics, McGill University, 3600 rue University, Montreal, Quebec, Canada H3A 2T8}

\begin{abstract}
  Low-energy excitations in graphene exhibit relativistic properties
  due to the linear dispersion relation close to the Dirac points in
  the first Brillouin zone. Two of the Dirac points located at
  opposite corners of the first Brillouin zone can be chosen as
  inequivalent, representing a new \textit{valley} degree of freedom,
  in addition to the charge and spin of an electron. Using the valley
  degree of freedom to encode information has attracted significant
  interest, both theoretically and experimentally, and gave rise to
  the field of \textit{valleytronics}. We study a graphene $p$-$n$
  junction in a uniform out-of-plane magnetic field as a platform to
  generate and controllably manipulate the valley polarization of
  electrons. We
  show that by tuning the external potential giving rise to the
  $p$-$n$ junction we can switch the current from one valley
  polarization to the other. We also consider the effect of
  different types of edge terminations and present a setup, where we
  can partition an incoming valley-unpolarized current into two
  branches of valley-polarized currents. The branching ratio can be
  chosen by changing the location of the $p$-$n$ junction using a
  gate.
\end{abstract}

\date{\today}
\pacs{81.05.ue, 73.40.-c, 73.43.-f}
\maketitle

\section{Introduction}
Two-dimensional (2D) materials are promising candidates for future
electronics due to their unique characteristics.  The pioneering 2D
material, graphene, was experimentally isolated in
2004~\cite{Novoselov2004}. The bandstructure of $p_z$ electrons in
single-layer graphene, modeled as a honeycomb lattice with lattice
constant $a=0.246$\:nm consisting of two triangular Bravais sublattices $A$ and
$B$ with nearest-neighbor hopping in the tight-binding formulation,
hosts six Dirac cones resulting from touching of the valence and
conduction bands at the Fermi energy $E=0$. Two of the cones located
at diagonally opposite corners of the first Brillouin zone can be
chosen as inequivalent, for example at $K=2\pi/3a$ and $-K$. For the
low-energy electronic excitations in the system they represent a new
degree of freedom of an electron, in addition to the charge and
spin. This \textit{valley} degree of freedom can be exploited in
analogy with the spin in spintronics, which gave rise to the field
called \textit{valleytronics}, where one uses the valley degree of
freedom to encode information.

There is a strong motivation to generate, controllably manipulate and
read out states of definite valley polarization, and a substantial
amount of theoretical and experimental work has been done towards
achieving these goals. A recent review of some advances made in the
field of valleytronics in 2D materials is provided in
Ref.~\onlinecite{Schaibley2016}. To mention some: a gated graphene
quantum point contact with zigzag edges was proposed to function as a
valley filter~\cite{Rycerz2007}. Superconducting contacts were shown
to enable the detection of the valley polarization in
graphene~\cite{Akhmerov2007}.  In 2D honeycomb lattices with broken
inversion symmetry, e.g. transition metal dichalcogenide (TMD)
monolayers, a non-zero Berry curvature carries opposite signs in the
$K$ and $-K$ valleys.  In these 2D materials, the velocity in the
direction perpendicular to an applied in-plane electric field is
proportional to this Berry curvature~\cite{Xiao2010}.  Hence the
electrons acquire a valley-antisymmetric transverse velocity leading to the
valley Hall effect, which spatially separates different valley states.
In a system where the occupation numbers of the two valleys are
different (valley-polarized system), a finite transverse voltage
across the sample is developed and the sign of this voltage can be
used to measure the valley polarization~\cite{Xiao2007}. The valley
Hall effect can also be exploited in a biased bilayer graphene, where
the out-of-plane electric field breaks the inversion
symmetry~\cite{Li2016,Sui2015,Shimazaki2015}. Moreover, it was shown
that the broken inversion symmetry results in the valley-dependent
optical selection rule, which can be used to selectively excite
carriers in the $K$ or $-K$ valley via right or left circularly
polarized light, respectively~\cite{Yao2008,Xiao2012}. Valley polarization can also be achieved in monolayer\cite{Pereira2009,Masir2011,Moldovan2012,Cheng2016} and bilayer\cite{Cheng2016} graphene systems with barriers. In addition, proposals exploiting strain that induces pseudomagnetic fields acting oppositely in the two valleys\cite{Milovanovic2016,Settnes2016} together with artificially induced carrier mass and spin-orbit coupling\cite{Grujic2014} have been put forward.

In this paper we propose a way to generate and controllably manipulate
the valley polarization of electrons in a graphene $p$-$n$ junction in
the presence of an out-of-plane magnetic field. Applying an
out-of-plane magnetic field to the graphene sheet leads to the
formation of low-energy relativistic Landau levels
(LLs)~\cite{Goerbig2011}. These are responsible for the unusual
quantum Hall conductance quantization
$G_n= 2_s \times 2_v \times(n+1/2)e^2/h$, where the integer $n$ is the
highest occupied Landau level index (for $n$-type doping) for a given
chemical potential. The factor $2_s$ in the formula accounts for the
spin degeneracy and the second factor $2_v$ for the valley degeneracy
of the Landau levels. The presence of the $E=0$ Dirac point and particle-hole
symmetry lead to a special $n=0$ LL, which is responsible for the
fraction $1/2$ in the conductance.

Semiclassically, charged particles propagating in a spatially varying
out-of-plane magnetic field in 2D may exhibit snake-like trajectories that
are oriented perpendicularly to the field gradient \cite{Mueller1992}.
The simplest case occurs along a nodal line of a spatially varying
magnetic field \cite{Park2008,Oroszlany2008,Ghosh2008,Prada2010}.
Another system, a graphene $p$-$n$ junction in a homogeneous
out-of-plane magnetic field, hosts similar states located at the
interface between $n$- and $p$-doped regions. These interface states
are also called snake states due to the shape of their semiclassical
trajectories~\cite{Carmier2010, Carmier2011, Rickhaus2015}. A
correspondence between these two kinds of snake trajectories was
pointed out in Ref.~\onlinecite{Beenakker2008a}. A mapping between
these two systems was found by rewriting both problems in a Nambu
(doubled) formulation~\cite{Liu2015}. In this paper we consider a
graphene $p$-$n$ junction in a homogeneous out-of-plane magnetic
field, a system which has attracted a lot of
attention~\cite{Cavalcante2016,Cohnitz2016,Fraessdorf2016,Kolasinski2016,Taychatanapat2015,Williams2007,Zarenia2013}.
In the limit of a large junction (where the phase coherence is
suppressed due to inelastic scattering or random time-dependent
electric fields), the conductance is a series conductance of $n$- and
$p$-doped regions~\cite{Abanin2007}. However, for sufficiently small
junctions the conductance depends on the microscopic edge termination
close to the $p$-$n$ interface. When the chemical potential in the $n$
and $p$ regions is within the first Landau gap, i.e., is restricted
to energy values smaller than the absolute value of the energy
difference between the zeroth and the first Landau level, an
analytical formula for the conductance can be
derived~\cite{Tworzydlo2007}, see Eq.~\eqref{eq:T_20}.

We demonstrate that a three-terminal device like the one shown in
Fig.~\ref{fig:1} can be used as a switchable, i.e. voltage-tunable
valley filter. In short, it works as follows: valley-unpolarized
electrons injected from the upper lead are collected in the lower
leads with high valley polarization.  The valley polarization of the
collected electrons is controlled by switching the $p$-$n$ junction on
and off, while the partitioning of the electron density between the
two lower leads is controlled by the edge termination and the width
$W$ of the central region close to the $p$-$n$ interface.

Our results are not restricted to graphene. They apply also to
honeycomb lattices with broken inversion symmetry, where the inversion
symmetry breaking term is represented by a staggered sublattice
potential. As long as the amplitude of this term is smaller than the
built-in potential step in the $p$-$n$ junction, our results remain
valid.  In a system with broken inversion symmetry, a non-zero Berry
curvature would give rise to the valley Hall effect which could be
used to read out the polarization of the outgoing
states~\cite{Xiao2007}.

The rest of this article is organized as follows. In Sec. II we
describe the setup of the proposed switchable valley filter and the
methods we use to investigate its properties in detail. In Sec. III we
present our numerical results, which demonstrate the valley-polarized
electronic transport. In Sec. IV we study the effect of potential step height and different edge terminations of the graphene lattice close to the $p$-$n$ interface on the valley polarization. We show that using a tilted staircase edge
$p$-$n$ junction allows to partition a valley-unpolarized incoming
current into two outgoing currents with opposite valley polarizations,
where the partitioning can be controlled by tuning the location of the
$p$-$n$ junction.  Finally we summarize our results in Sec. V.

\section{Setup}
\begin{figure}[h]
\includegraphics[width=0.99\columnwidth]{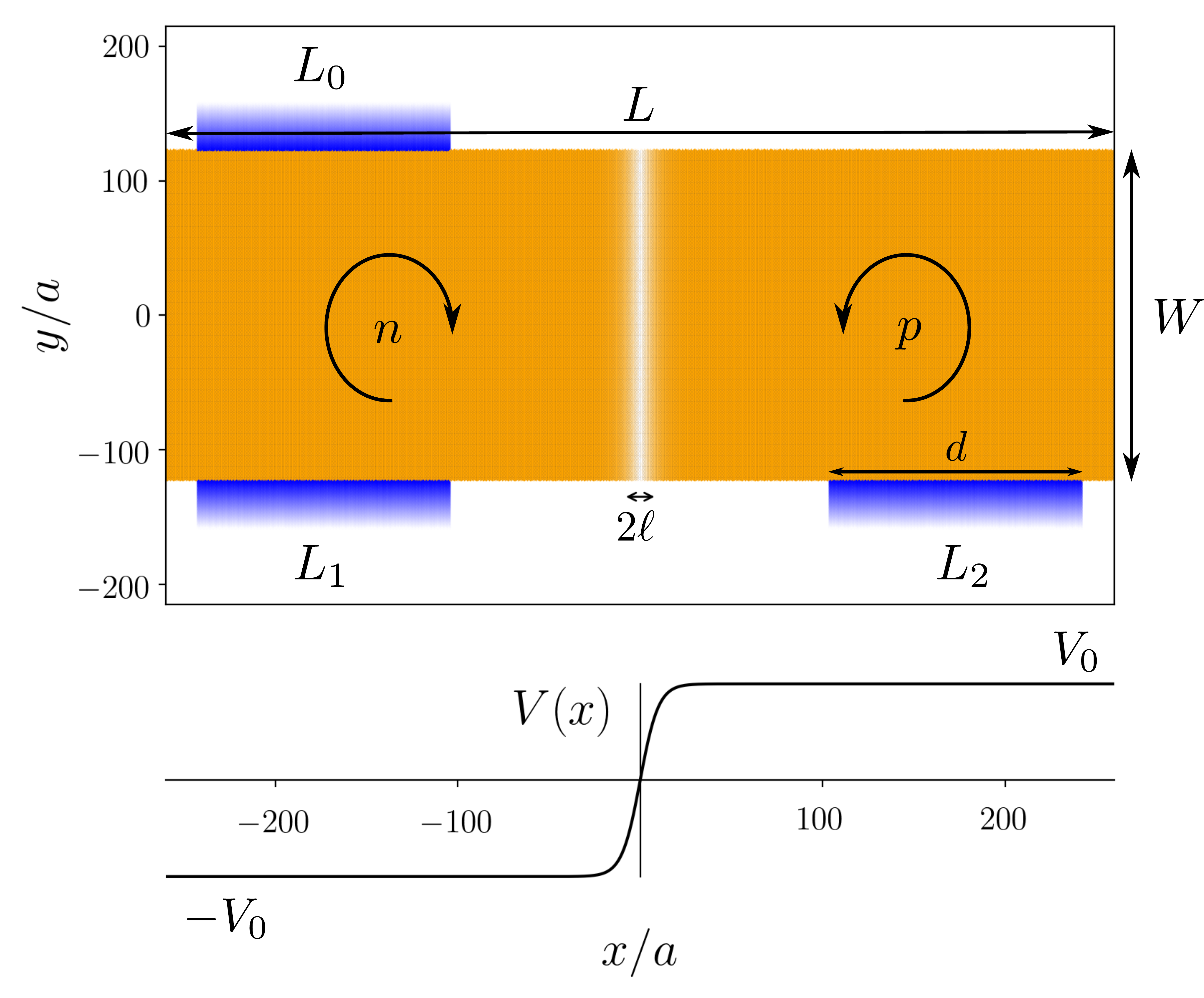}
\caption{Three-terminal device used as a switchable valley filter. The
  three leads of horizontal size $d$ with zigzag edge termination,
  upper lead $L_0$, lower-left lead $L_1$ and lower-right lead $L_2$,
  are attached to the rectangular scattering region of length $L$ and
  width $W$. Top and bottom edges of the scattering region are of
  armchair type, while the left and right edges are of zigzag
  type. The $p$-$n$ interface of thickness $2\ell$ (white color
  gradient) is modeled by an $x$-dependent on-site potential
  $V(x)$. In an out-of-plane magnetic field with $n$ Landau levels
  occupied ($n=0,1,2\dots$), there are $2n+1$ edge states with
  opposite chirality in the $n$- and $p$-doped region. Along the
  $p$-$n$ interface there are $2(2n+1)$ co-propagating snake states.}
\label{fig:1}
\end{figure}

Figure~\ref{fig:1} shows the three-terminal device that we would
like to study. A rectangular region of width $W$ and length $L$
represents the graphene $p$-$n$ junction in a uniform out-of-plane
magnetic field, also referred to as the scattering region. It is described
by a tight-binding Hamiltonian of the form
\begin{align}
  H = \sum_i \, V(\textbf{r}_i) c^\dag_i c_i +
  \sum_{\langle i,j \rangle} \, t e^{i \varphi_{ij}} c^\dag_i c_j\:,
\label{eq:tb_hamiltonian}
\end{align}
where $V(\textbf{r}_i)$
is the scalar on-site potential at site $i$ with
coordinate $\mathbf{r}_i$ and $\varphi_{ij} = (e/\hbar) \int_i^j
\mathbf{A} \cdot \mathrm{d} \mathbf{r}$ is the Peierls phase accumulated along
the link from site $i$ to site $j$ in magnetic field $\mathbf{B} =
[0,0,B]$. The Zeeman splitting is neglected, i.e., we consider
spinless electrons. The sum over $\langle i,j \rangle$ denotes the sum
over nearest neighbors.
We choose the Landau gauge, where the vector potential is
\begin{equation*}
\mathbf{A}=[0,Bx,0]\:.
\end{equation*}
In this gauge we can define a (quasi-)momentum parallel to the edges
of the leads. The leads $L_0$, $L_1$, and $L_2$ are modeled as
semi-infinite zigzag nanoribbons, where the valley index can be well
distinguished in $k$-space. They are also described by the Hamiltonian
\eqref{eq:tb_hamiltonian} and below we present the case with $B=0$ in
the leads, which is however not crucial for our results. The Peierls
phase $\varphi_{ij}$ can be written in the form
\begin{equation*}
  \varphi_{ij}= 2\pi\frac{\phi}{\phi_0}\frac{2}{\sqrt{3}a^2}
  \frac{x_i+x_j}{2}(y_j-y_i)\:,
\end{equation*}
where $\phi_0=h/e$ is the magnetic flux quantum and $\phi=B S$ is the
flux through a single hexagonal plaquette of a honeycomb
lattice. Here, $S=\sqrt{3}a^2/2$ is the area of a hexagonal plaquette
of the honeycomb lattice. An important length scale derived from
magnetic field is the magnetic length $\ell_B=\sqrt{\hbar/eB}$. In the
rest of the paper the magnetic field is chosen such that
$\phi/\phi_0=0.003$ and hence $\ell_B\approx 6.78a$. We also denote
the energy difference between the first and the zeroth LL by
$\delta=\sqrt{2}\hbar v_F/\ell_B=0.18t$, where the Fermi velocity at
the $K$ point is $v_F=\sqrt{3}at/2\hbar$.

The scalar on-site potential is varying only in the $x$-direction as
\begin{equation*} \label{eq:on-site_potential}
 V(x) = V_0 \tanh(x/\ell)\:,
\end{equation*}
where $V_0$ is the external scalar potential and $2\ell$ is the
thickness of the domain wall characterizing the $p$-$n$ junction.  If
both $V_0$ and $B$ are non-zero and $V_0<\delta$ in such a setup,
there exist two snake states co-propagating along the $p$-$n$
junction~\cite{Liu2015}. The orientation of the fields in the system
is such that the snake states are traveling downwards in the negative
$y$-direction.

We calculate the transmission $T_{10}$ from $L_0$ to $L_1$, and
$T_{20}$ from $L_0$ to $L_2$. We also calculate the valley-resolved
transmissions, with the following notation: $T_{20}^{K(-K)}$ is the
transmission from $L_0$ to $L_2$, where in $L_2$ we sum only over
outgoing modes with $k\in (0,\pi/a]$ ($k\in (-\pi/a,0]$). Then we can
define the valley polarization in $L_2$ as
$P_2=(T_{20}^K-T_{20}^{-K})/T_{20}$.
Analogous quantities are defined for $L_1$.

Due to the absence of backscattering in chiral quantum Hall edge
states and the symmetry of the $p$-$n$ junction for $E=0$, the net
transmission (no spin) is $T_{10}+T_{20}=2n+1$, where $n$ is the
highest occupied LL in the $n$-doped region. The partitioning of the
net transmission between $T_{10}$ and $T_{20}$ depends on the edge
termination close to the $p$-$n$ interface according to the
formula~\cite{Tworzydlo2007}
\begin{equation}
 T_{20}=\frac{1}{2}(1-\cos\Phi)\:,
 \label{eq:T_20}
\end{equation}
where $\Phi$ is the angle between valley isospins at the upper and
lower edge represented as vectors on the Bloch sphere. For armchair
edges one has $\Phi=\pi$ if $W/a \mod 3=0$ and $\Phi=\pm\pi/3$
otherwise. The formula is valid if the $n$ and $p$ regions are on the
lowest Hall plateau, where the quantum Hall conductances in the $n$-
and $p$-doped regions are equal to $e^2/h$ (ignoring the spin degree
of freedom)~\cite{Tworzydlo2007}.  The transmission from $L_0$ to
$L_1$ is then given by $T_{10}=1-T_{20}$. Interference between
wavefunctions of the snake states is responsible for this
partitioning. The snake-state wavefunctions are located at the $p$-$n$
interface and their effective spread in the $x$-direction is given by
the magnetic length $\ell_B$ to the left and right of the
interface. Hence one way to control the partitioning experimentally
will be to control the edge termination around the $p$-$n$ interface on
a length scale of the order of $2\ell_B$.

In the following we show numerically that by switching the $p$-$n$
junction on ($V_0\neq 0$) and off ($V_0=0$) we can control the valley
polarization of the outgoing states in leads $L_1$ and $L_2$.

\section{Switchable valley filter}
We demonstrate the principle of the switchable valley filter using a
$p$-$n$ junction in an out-of-plane magnetic field, where the upper
and the lower edges are of armchair type. The system is described by
the Hamiltonian shown in Eq.~\eqref{eq:tb_hamiltonian}. The width $W$
of the $p$-$n$ junction is chosen such that the number of hexagons in
this width is a multiple of $3$, i.e.  such that the corresponding
armchair nanoribbon would be metallic. If, furthermore, both the $n$-
and $p$-doped regions are on the lowest Hall plateau and $V_0$ is
large enough~\cite{Tworzydlo2007}, we expect $T_{20}=1$ and
$T_{10}=0$, see Eq.~\eqref{eq:T_20}. The switchable valley filter is
based on the fact that for a zigzag graphene nanoribbon the quantum
Hall edge states of the $n=0$ LL lying in opposite valleys $K$ and $-K$
have opposite velocities~\cite{Brey2006,Beenakker2008b,Goerbig2011}.

Unless stated otherwise, the system has length $L=520a$ and width
$W=246a$ (the exception is Fig.~\ref{fig:4}). The horizontal size of
each lead is $d=156a$. The thickness of the $p$-$n$ junction is
$2\ell=20a$. We set the magnitude of the magnetic field in the leads
to zero, which is however not crucial for the result. Our
tight-binding calculations were performed using
Kwant~\cite{Groth2014}.

\begin{figure}[h]
\centering
\subfloat[]{\includegraphics[width=0.95\columnwidth]{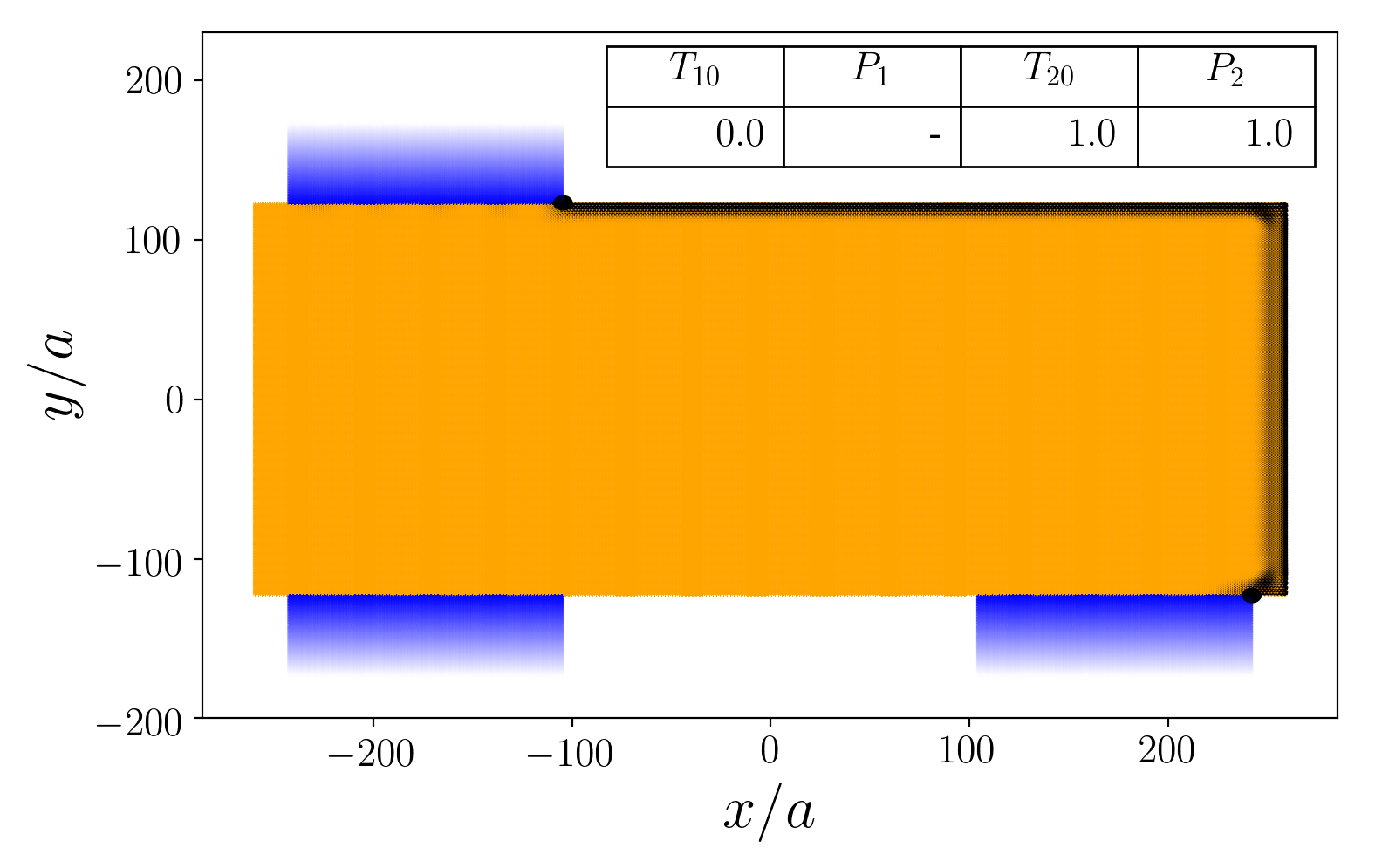}}\\
\subfloat[
]{\includegraphics[width=0.95\columnwidth]{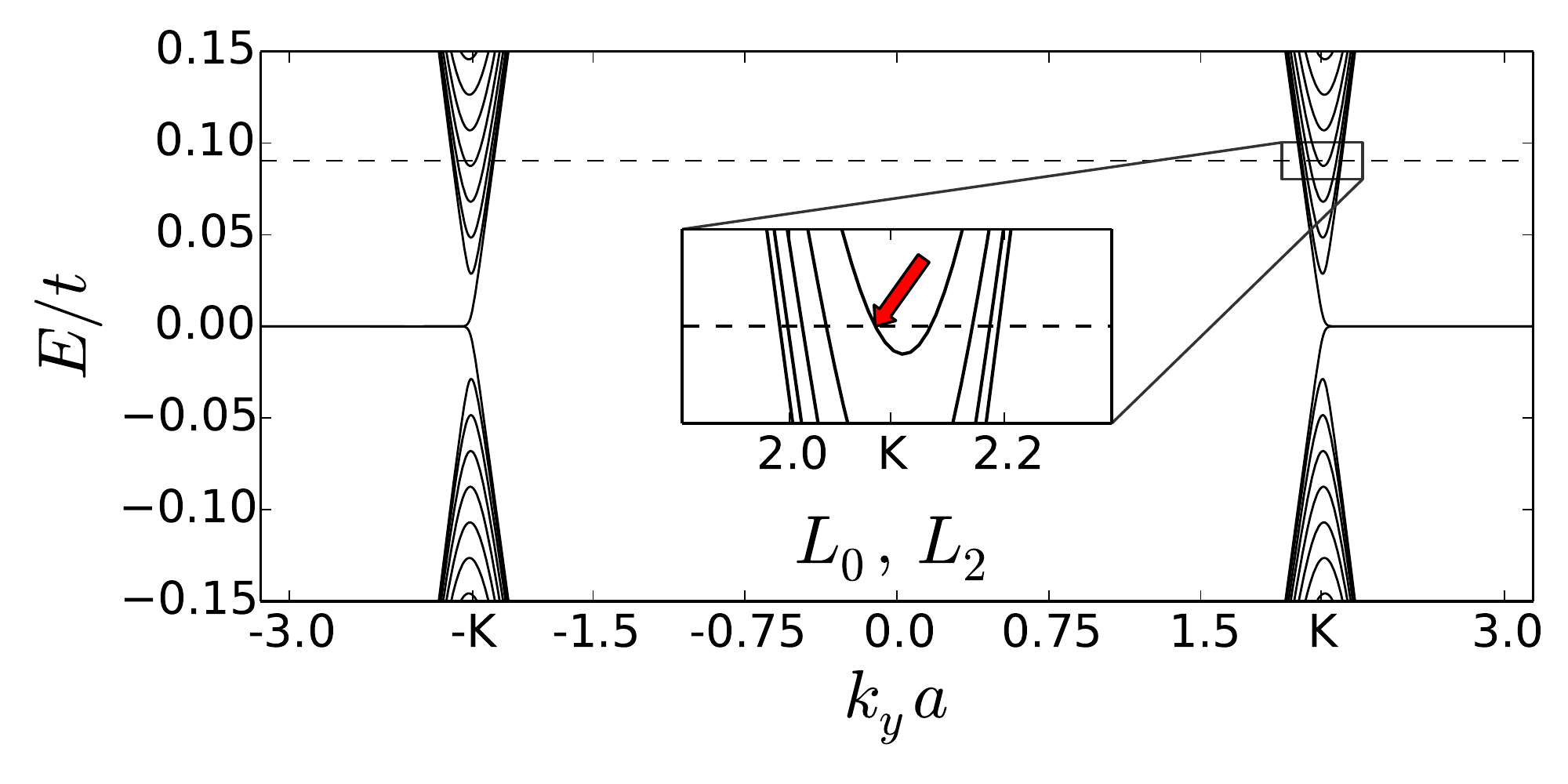}}
\caption{Case when $V_0=0$ and $E=\delta/2$. (a) State in the
  scattering region due to incoming mode from lead $L_0$ with
  $k_y a = 2.08$. The inset table lists the transmissions $T_{10}$
  and $T_{20}$ and polarizations $P_1$ and $P_2$. (b) Bandstructure of
  leads $L_0$ and $L_2$. The Fermi energy is indicated by the
  horizontal dashed line. The red arrow indicates the incoming mode on
  lead $L_0$ which has velocity $v<0$ and is chosen to be plotted in (a).}
\label{fig:2}
\end{figure}

\begin{figure}[h]
\centering
\subfloat[]{\includegraphics[width=0.95\columnwidth]{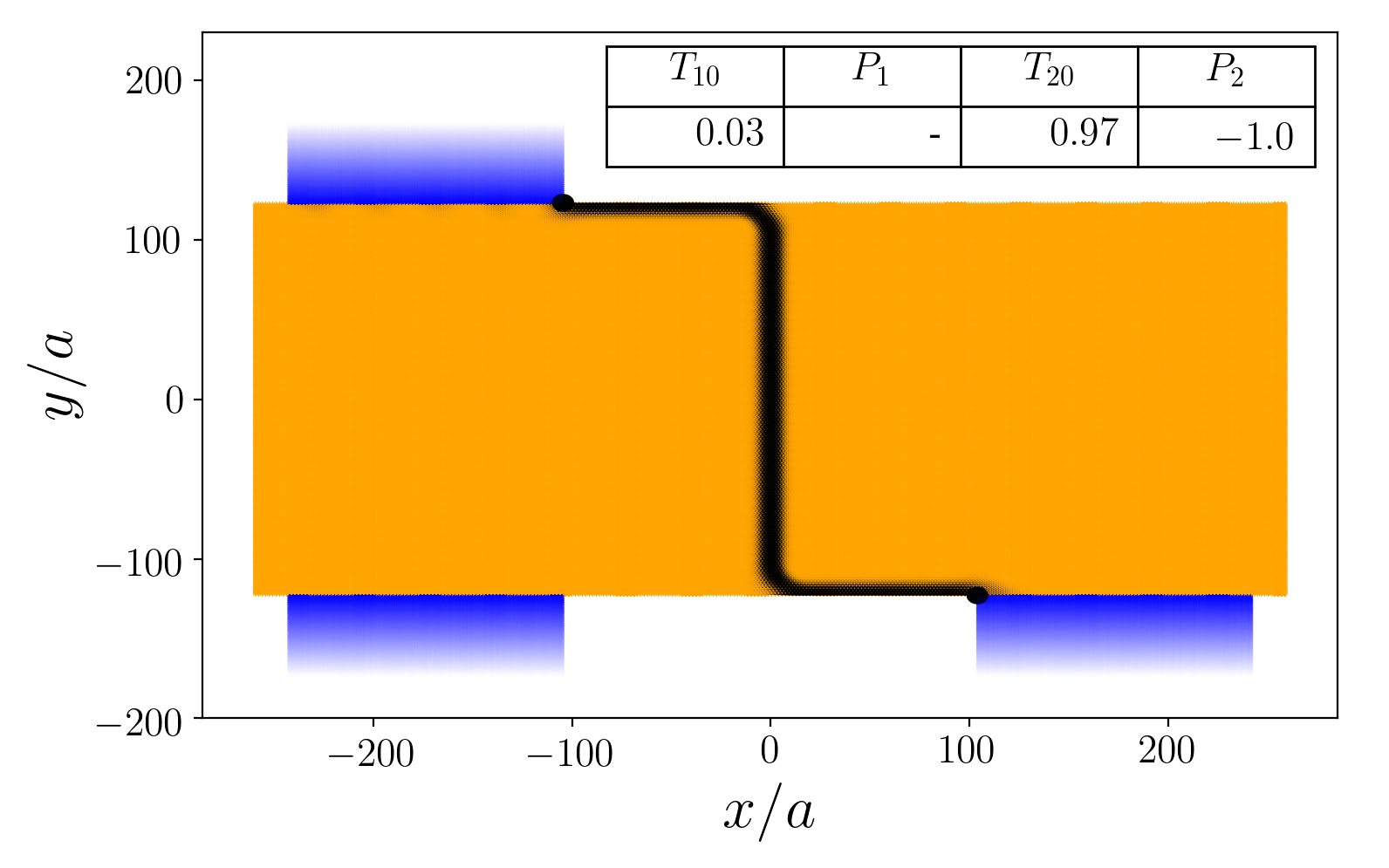}}\\
\subfloat[
]{\includegraphics[width=0.95\columnwidth]{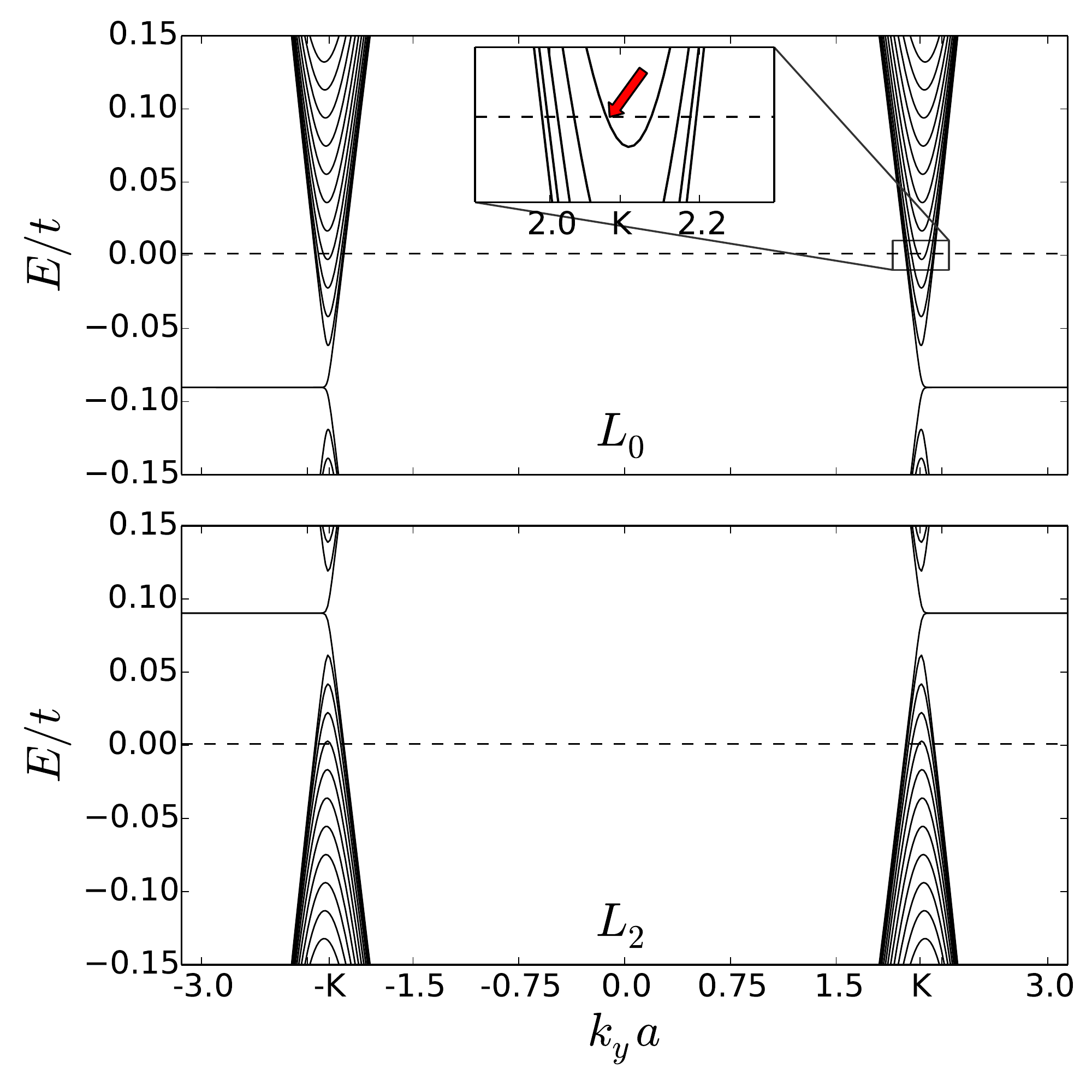}}
\caption{Case when $V_0=\delta/2$ and $E=0.001t$. (a) State in the
  scattering region due to incoming mode from lead $L_0$ with
  $k_y a =2.08$. The inset table lists the transmissions $T_{10}$ and
  $T_{20}$ and polarizations $P_1$ and $P_2$. (b) Top: bandstructure
  of lead $L_0$. The Fermi energy is indicated by the horizontal
  dashed line. The red arrow indicates the incoming mode chosen to
  plot (a). Bottom: bandstructure of lead $L_2$.}
\label{fig:3}
\end{figure}

First, we consider the case $V_0=0$.  A valley-unpolarized electron
current (injected from both valleys) in $L_0$ ends up as outgoing
valley-polarized electron current in $L_2$.  In Figs.~\ref{fig:2}(a)
and \ref{fig:3}(a) we also plot the probability density of one of the
states carrying the current by drawing a black dot on each site whose
size is proportional to the probability of finding an electron on that
particular site.  This is plotted for a state in the scattering region
due to an incoming mode from $L_0$ at Fermi energy $E$ and with
momentum $k_y$ indicated by the red arrow in the bandstructure for
$L_0$, see Figs.~\ref{fig:2}(b) and \ref{fig:3}(b).
Figure~\ref{fig:2}(a) shows the probability density of the state in
the scattering region due to an incoming mode from $L_0$ at
$E=\delta/2$ and $k_y a=2.08$. Since $V_0=0$, there are no snake
states in this system and the electronic current is carried by the
quantum Hall edge states.  The electrons injected from $L_0$ travel in
a clockwise manner to $L_2$. The calculated transmissions $T_{10}$ and
$T_{20}$ and polarizations $P_1$ and $P_2$ are shown in the inset
table in Fig.~\ref{fig:2}(a). We find that the outgoing electrons in
$L_2$ are perfectly polarized in the $K$ valley. This shows the
valley-polarized nature of the zeroth LL. Thus, this system can be
used as a valley filter.

If we now turn on the $p$-$n$ junction, the situation will change. We
assume that the $p$-$n$ junction is turned on adiabatically. We choose
$V_0=\delta/2$, so that the $n$- and $p$-doped regions are on the
$n=0$ LL. Figure~\ref{fig:3}(a) shows the probability density of the
state in the scattering region due to the incoming mode from $L_0$ at
the Fermi energy $E=0.001t$ and with $k_y a=2.08$ (red arrow in the
inset of Fig.~\ref{fig:3}(b)). In this system there are two
co-propagating snake states along the $p$-$n$ interface, and the
electronic current is carried by these states. These snake states are
located at the $p$-$n$ interface and spread in the $\pm x$-direction
over the magnetic length $l_B$, which is independent of the domain
wall thickness $2\ell$.

The electrons injected from $L_0$ now travel along the upper edge in
the $n$ region towards the $p$-$n$ interface and continue along the
$p$-$n$ interface towards the lower edge, where they enter the $p$
region with probability $\approx 1$ due to the specifically chosen $W$
and the armchair edge termination at both ends of the
interface. Finally, they end up in $L_2$. The corresponding
transmissions and polarizations are shown in the inset table of
Fig.~\ref{fig:3}(a). We find that the electrons in $L_2$ are nearly
perfectly polarized in the $-K$ valley. Thus, by turning on the $p$-$n$
junction we have \textit{flipped} the valley polarization of the
electronic current in $L_2$.

It is worth noting that these results are robust with respect to edge
disorder because of the absence of backscattering in the chiral
quantum Hall edge states. Our results also apply to the case where
the magnetic field is present in the leads. However, since for
low energies and dopings, only the $n=0$ LL plays a role, states in
the leads are already valley-polarized edge states and thus our
three-terminal device would then work as a perfect valley switch.

\section{Polarizations and transmissions upon varying $V_0$ and geometry}
In the following we analyze how the valley polarizations $P_1$ and $P_2$ and transmissions $T_{10}$ and $T_{20}$ change upon varying $V_0$ and edge terminations.

\begin{figure}[h]
\centering
\includegraphics[width=0.99\columnwidth]{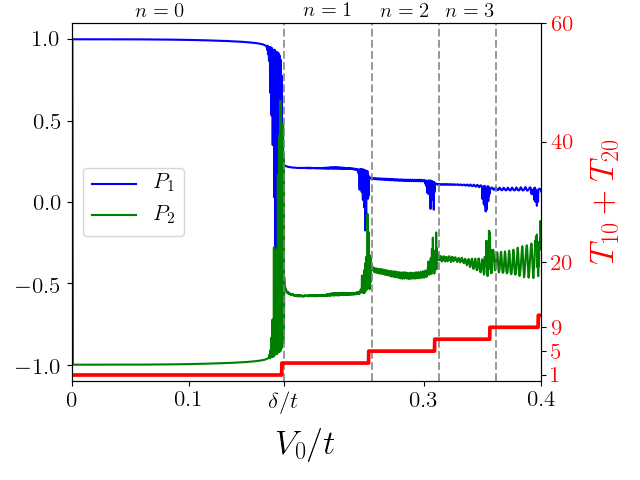}
\caption{Polarization $P_1$ in lead $L_1$ (blue, left axis), and $P_2$
  in lead $L_2$ (green, left axis) as a function of $V_0$ for the
  device shown in Fig.~\ref{fig:1}.  For $0<V_0<\delta$ only the
  $n=0$ Landau level (LL) is occupied and $P_2\approx -1$. As soon as
  higher LLs get involved ($V_0>\delta$), where the edge states are
  not valley-polarized, the valley polarization in $L_1$ and $L_2$
  decreases towards $0$ with increasing $V_0$.  The sum of the
  transmissions $T_{10}+T_{20}$ (red step-like curve, right axis)
  exhibits quantization due to LLs in the scattering region. The
  device is a good valley filter for $V_0<\delta$, i.e., when only the
  $n=0$ LL is occupied.
  Vertical (grey dashed) lines mark the LL energies
  $E_n=\sqrt{2n}\hbar v_F/\ell_B$ in the $n$-doped region calculated
  for a linear Dirac dispersion. The parameters chosen for this figure
  are $L=2080a$, $W=984a$, $\ell=40a$, and $d=320a$.}
\label{fig:4}
\end{figure}

{\it Polarization vs. $V_0$}. In Fig.~\ref{fig:4} we plot the
polarization in the leads as a function of $V_0$. Let us focus on
$P_2$, because the majority of electrons are traveling into lead $L_2$
(for this specifically chosen scattering region). The case shown in
Fig.~\ref{fig:2} corresponds to $V_0=0$, where $P_2=1$ (not visible in
the figure). For $V_0>0$ (Fig.~\ref{fig:3}), the polarization in lead
$L_2$ changes sign, $P_2=-1$. For $0<V_0<\delta=0.18t$ only the $n=0$
LL valley-polarized edge states contribute to the transport and $P_2$
stays close to $-1$ until $V_0\approx\delta$. For $V_0>\delta$ the
higher LLs get occupied. Edge states in the higher LLs are not
valley-polarized which reduces $P_2$. On further increasing $V_0$
higher and higher LLs get occupied which further obscures the edge
state valley polarization of the $n=0$ LL and the magnitude of $P_2$
decreases. The population of the LLs can be seen from $T_{10}+T_{20}$
as a function of $V_0$, which is shown with the red curve in
Fig.~\ref{fig:4}. Hence the efficiency of the switchable valley filter
device is observed to decrease with an increase in $V_0$, i.e. with
populating higher LLs.

\begin{figure}
\centering
\includegraphics[width=0.95\columnwidth]{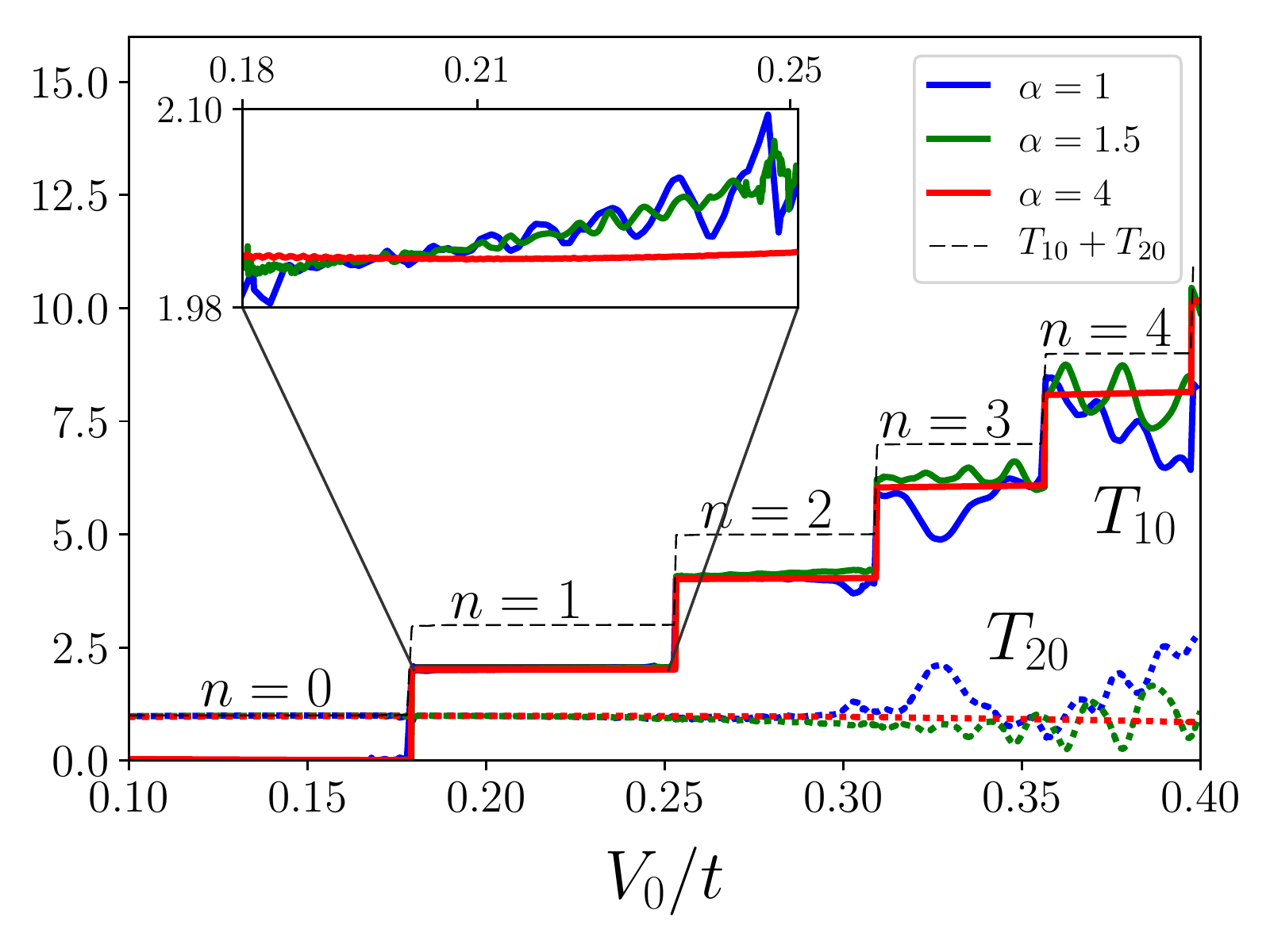}
 \caption{Transmissions $T_{10}$ (solid lines), $T_{20}$ (dotted lines) and $T_{10}+T_{20}$ (dashed line) as a function of $V_0$ for different system sizes. The inset shows that the amplitude of the oscillations tend to vanish as we increase the system size, which is accomplished by multiplying the parameters $L,W,d$ and $\ell$ by a factor of $\alpha=1$ (blue), $1.5$ (green), $4$ (red) while $\phi/\phi_0=0.003$ is kept constant.}
 \label{fig:5}
\end{figure}

Furthermore, one can notice in Fig.~\ref{fig:4} that the valley
polarizations $P_1$ and $P_2$ in leads $L_1$ and $L_2$ jump at the
same voltages $V_0$, where the $p$-$n$ junction undergoes quantum Hall
transitions and the total transmission $T_{10} +T_{20}$ changes by
$2$. The larger $V_0$, the more the values of $V_0$ at which the total
transmission changes by 2 deviate from the vertical lines
indicating the LL energies at $E_n=\sqrt{2n}\hbar v_F/\ell_B$.
This is due to the nonlinearity of the
dispersion which leads to a change in group velocity.
The oscillations seen in the blue and green curves stem from oscillations of transmissions as shown in Fig.~\ref{fig:5}. This can be viewed as a consequence of interference effects between modes confined at the $p$-$n$ interface with different momenta\cite{Kolasinski2016,Milovanovic2014}. However, the amplitude of these oscillations decreases with increasing system size. In our simulations, the system size is increased proportionally, i.e. parameters $L,W,d$ and $\ell$ are multiplied by $\alpha = 1, 1.5, 4$, while the magnetic flux per plaquette is kept constant, $\phi/\phi_0 = 0.003$ (see Fig.~\ref{fig:5}).


{\it Different edge terminations.} We find that different edge
terminations and $p$-$n$ interface length have almost no influence on
the valley polarization in the leads, but they determine the
partitioning of the net transmission between $T_{10}$ and
$T_{20}$. In Tab.~\ref{tab:additional_cases}(a), where the $p$-$n$ interface meets armchair edges, $T_{10}$ and $T_{20}$ exhibit the expected periodicity when changing the width $W$ such that the number of hexagons across the width of the scattering region changes by $3$. The case in which the $p$-$n$ interface meets zigzag edges is considered in Tab.~\ref{tab:additional_cases}(b). Here, the transmissions $T_{10}$ and $T_{20}$ switch values depending on whether the two edges are in zigzag or anti-zigzag configuration, which is in agreement with Ref.~\onlinecite{Akhmerov2008}. To model different edge terminations, we also added a triangular region to the sample, see Tab.~\ref{tab:additional_cases}(c)--(e) (a zoom-in onto the tip is shown in the last column of
Tab.~\ref{tab:additional_cases}(c)).   
Thus by controlling the edge termination on a length scale of
$2\ell_B$ around the $p$-$n$ interface one can tune the partitioning of
the current into $L_1$ and $L_2$. The currents in both of these leads
are polarized in opposite valleys. Thus if one chooses the situation
where the current is finite in both $L_1$ and $L_2$ (for example the
case shown in Tab.~\ref{tab:additional_cases}(c)), one can create two
streams of oppositely valley-polarized currents in leads $L_1$ and
$L_2$.

\setlength{\extrarowheight}{0.6cm}
\setlength{\tabcolsep}{0.1cm}

\begin{table}[h]
     \begin{center}
     \begin{tabular}{ c | c | c c c c | c }
      & geometry & $T_{10}$ & $P_1$ & $T_{20}$ & $P_2$ & notes\\ 
     \hline
     (a) & \raisebox{-\totalheight/2}{\includegraphics[width=0.24\columnwidth, height=0.15\columnwidth]{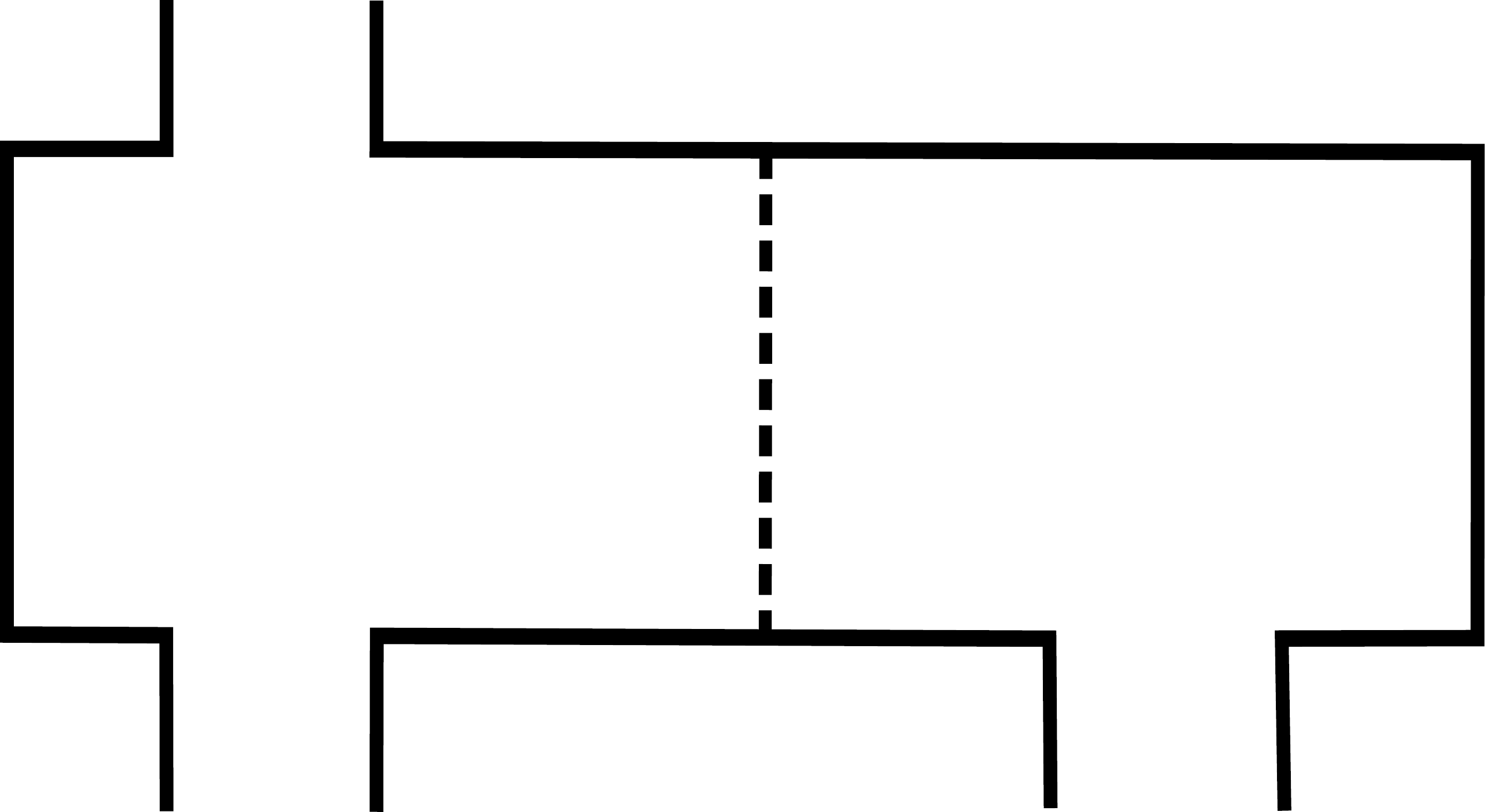}}
      &  \begin{tabular}{@{}c@{}} 0.03 \\ 0.6 \\ 0.88 \end{tabular}  &  \begin{tabular}{@{}c@{}} -- \\ 0.99 \\ 1 \end{tabular} &  \begin{tabular}{@{}c@{}} 0.97 \\ 0.4 \\ 0.12 \end{tabular}  &  \begin{tabular}{@{}c@{}} -1 \\ -0.99 \\ -1 \end{tabular} & \begin{tabular}{@{}c@{}} $W=$246a \\ $W=$247a \\ $W=$248a \end{tabular} 
            \\[0.7cm] \hline
	(b) & \raisebox{-\totalheight/2}{\includegraphics[width=0.24\columnwidth, 				height=0.15\columnwidth]{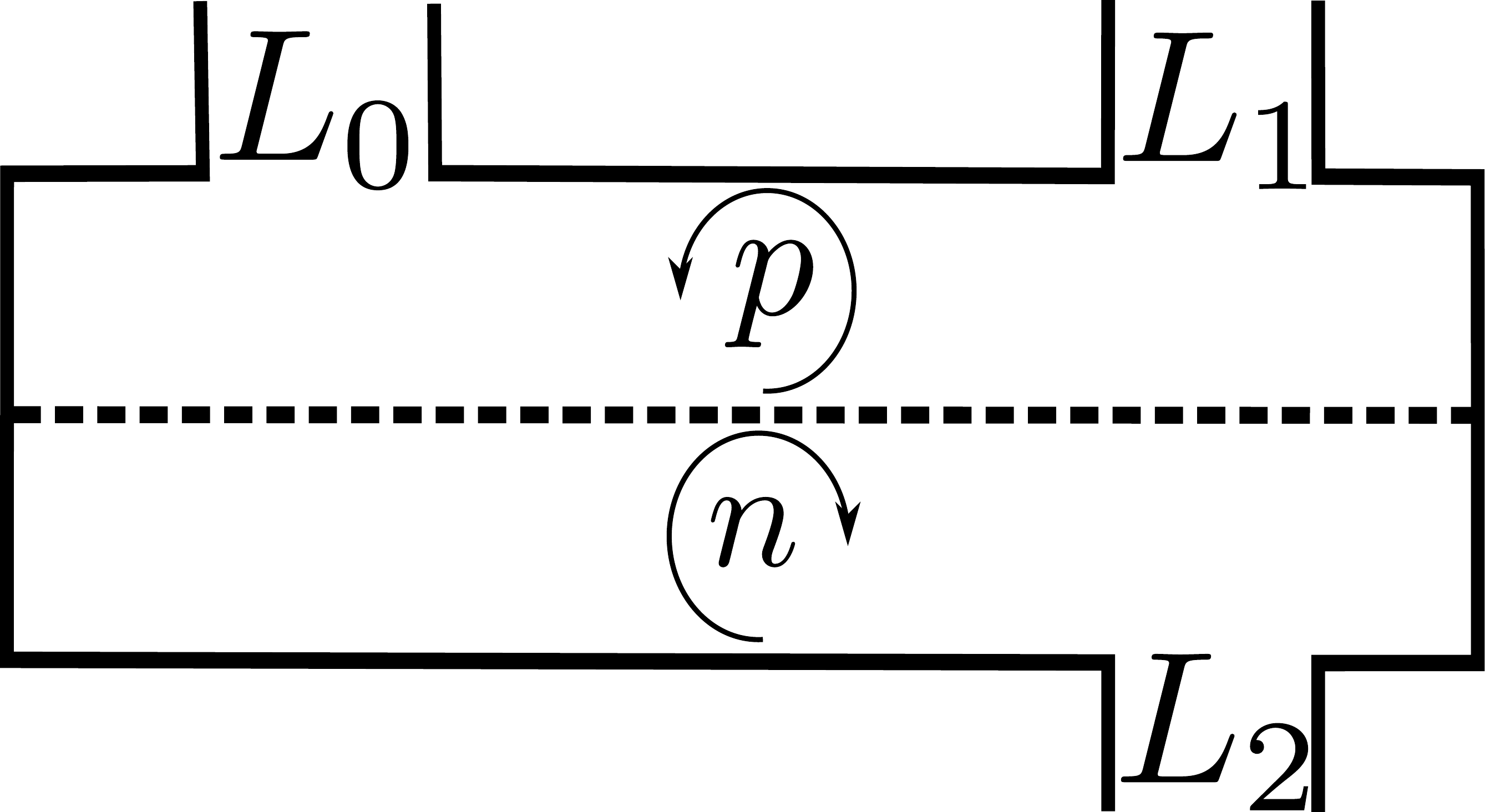}}
      &\begin{tabular}{@{}c@{}} 0.0 \\ 1.0  \end{tabular}  &  \begin{tabular}{@{}c@{}} -- \\ 0.93 \end{tabular} &  \begin{tabular}{@{}c@{}} 1.0 \\ 0.0 \end{tabular}  &  \begin{tabular}{@{}c@{}} -0.93 \\ -- \end{tabular} & \begin{tabular}{@{}c@{}} $L=$520a \\ $L=$521a \end{tabular}
      \\[0.7cm] \hline 
     (c) & \raisebox{-\totalheight/2}{\includegraphics[width=0.24\columnwidth, height=0.15\columnwidth]{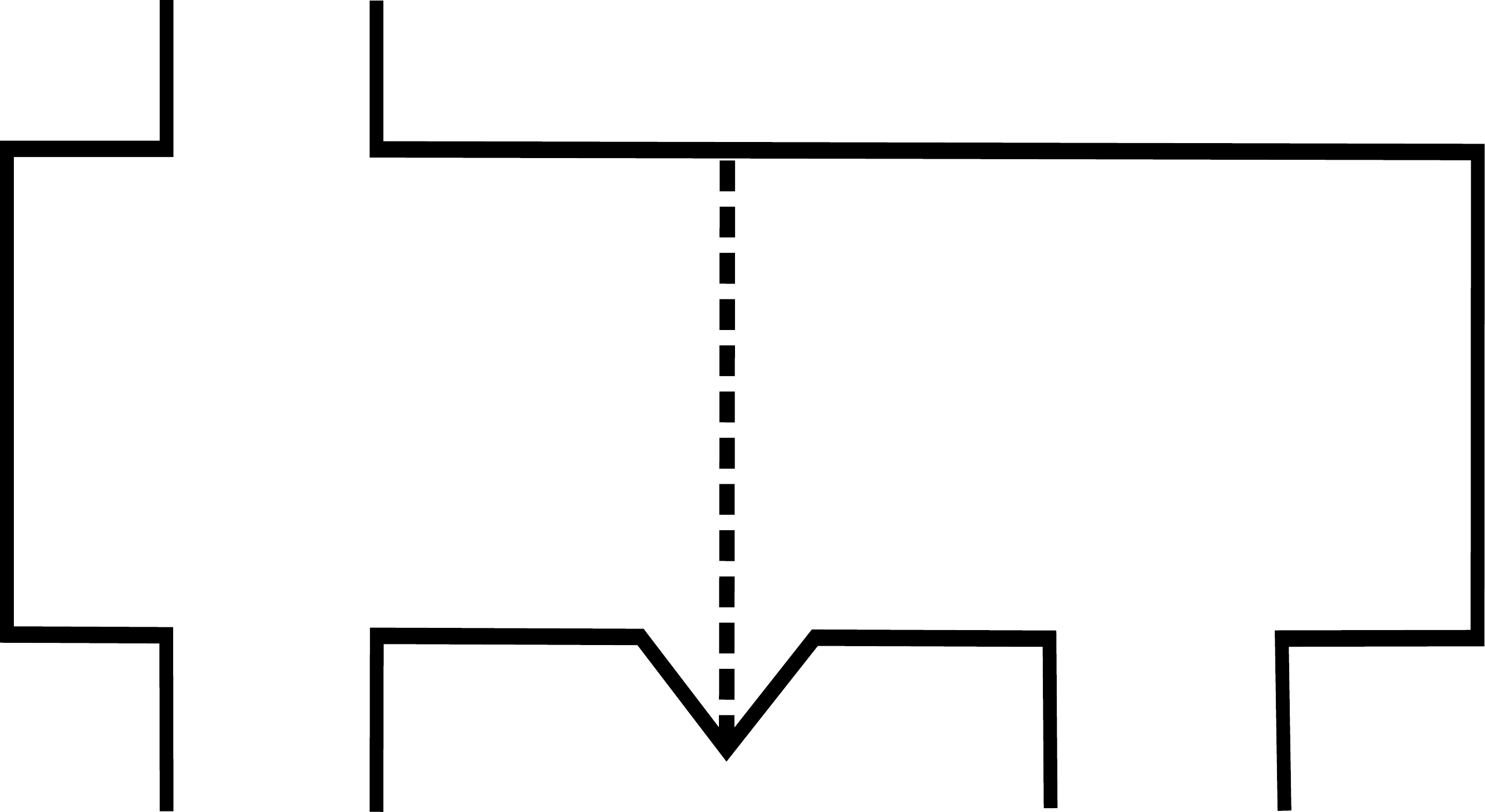}}
      &  0.48 &  1 &  0.52  &  -1  &  \raisebox{-\totalheight/2}{\includegraphics[width=0.22\columnwidth, height=0.1\columnwidth]{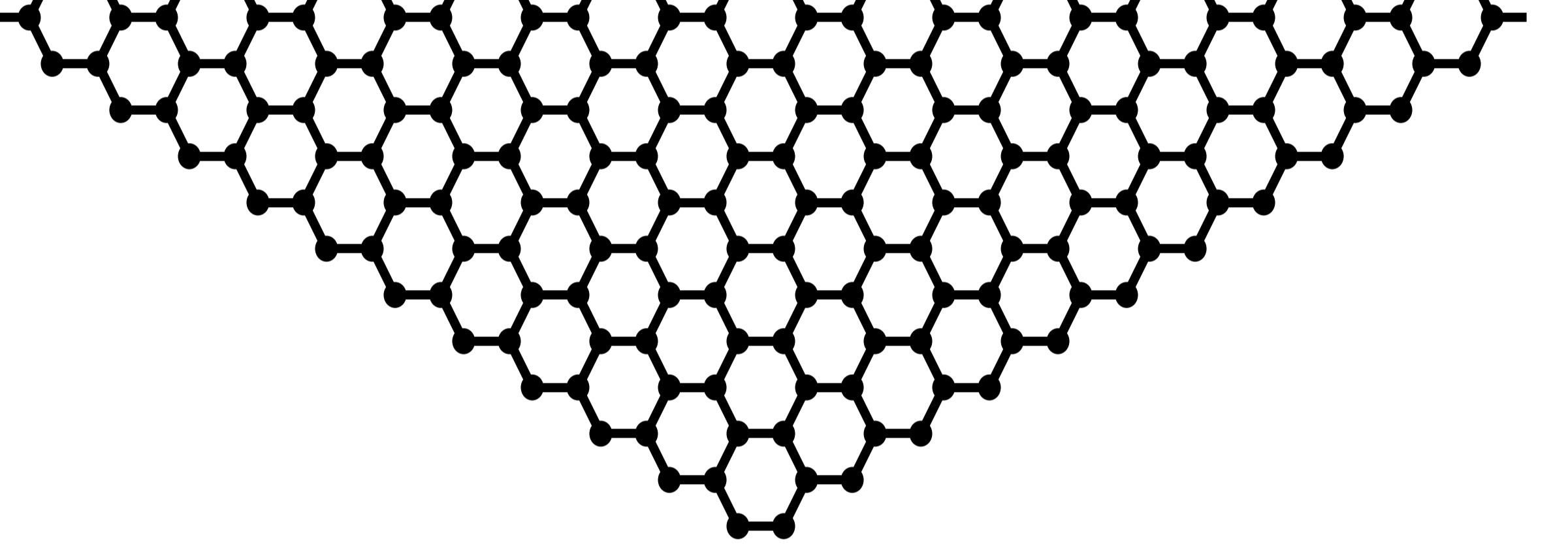}}  
      \\[0.7cm] \hline

     (d) & \raisebox{-\totalheight/2}{\includegraphics[width=0.24\columnwidth, height=0.15\columnwidth]{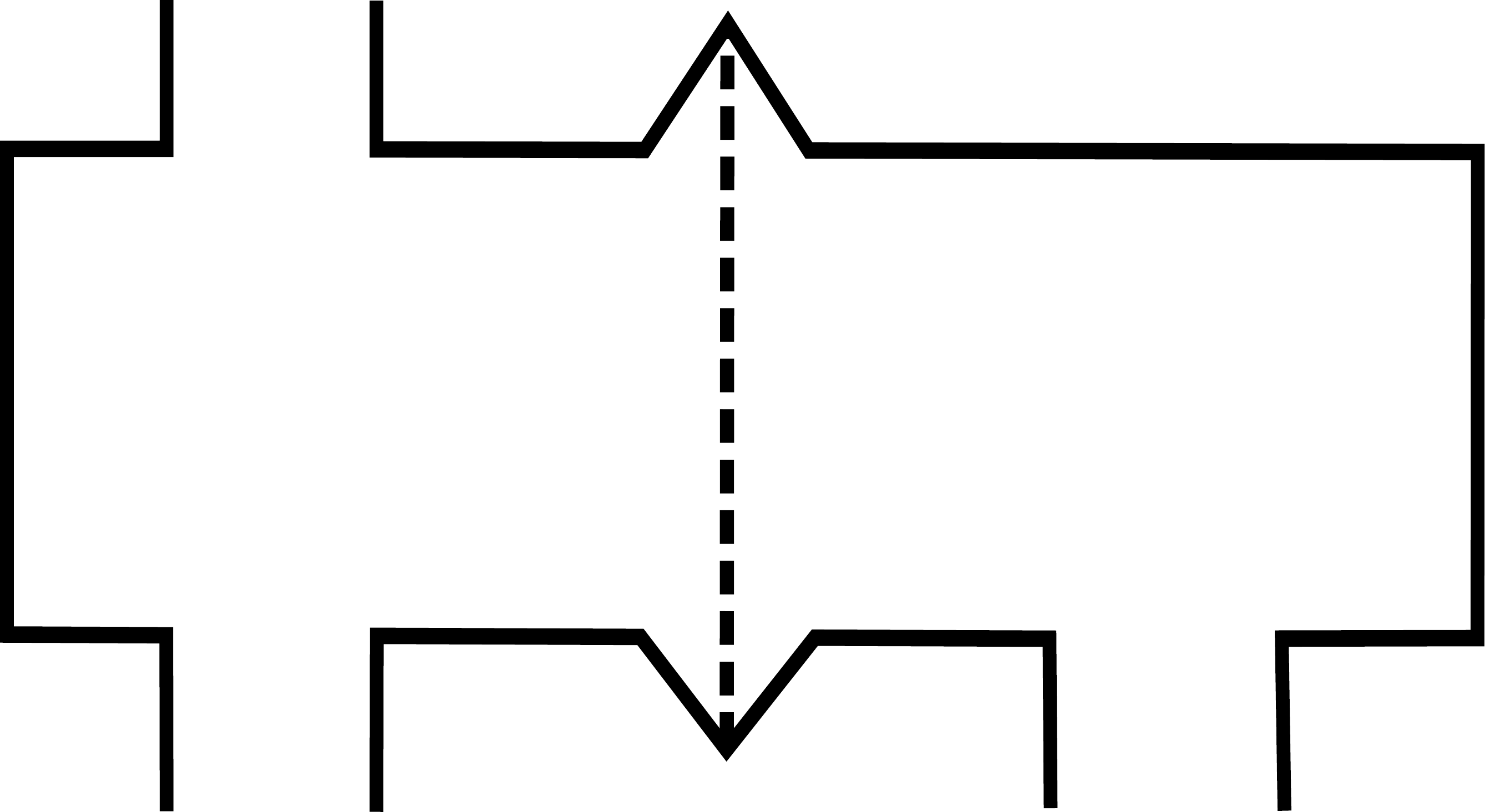}}
      & 1  & 1 &  0  &  --  &  W independent
            \\[0.7cm] \hline
     (e) & \raisebox{-\totalheight/2}{\includegraphics[width=0.24\columnwidth, height=0.15\columnwidth]{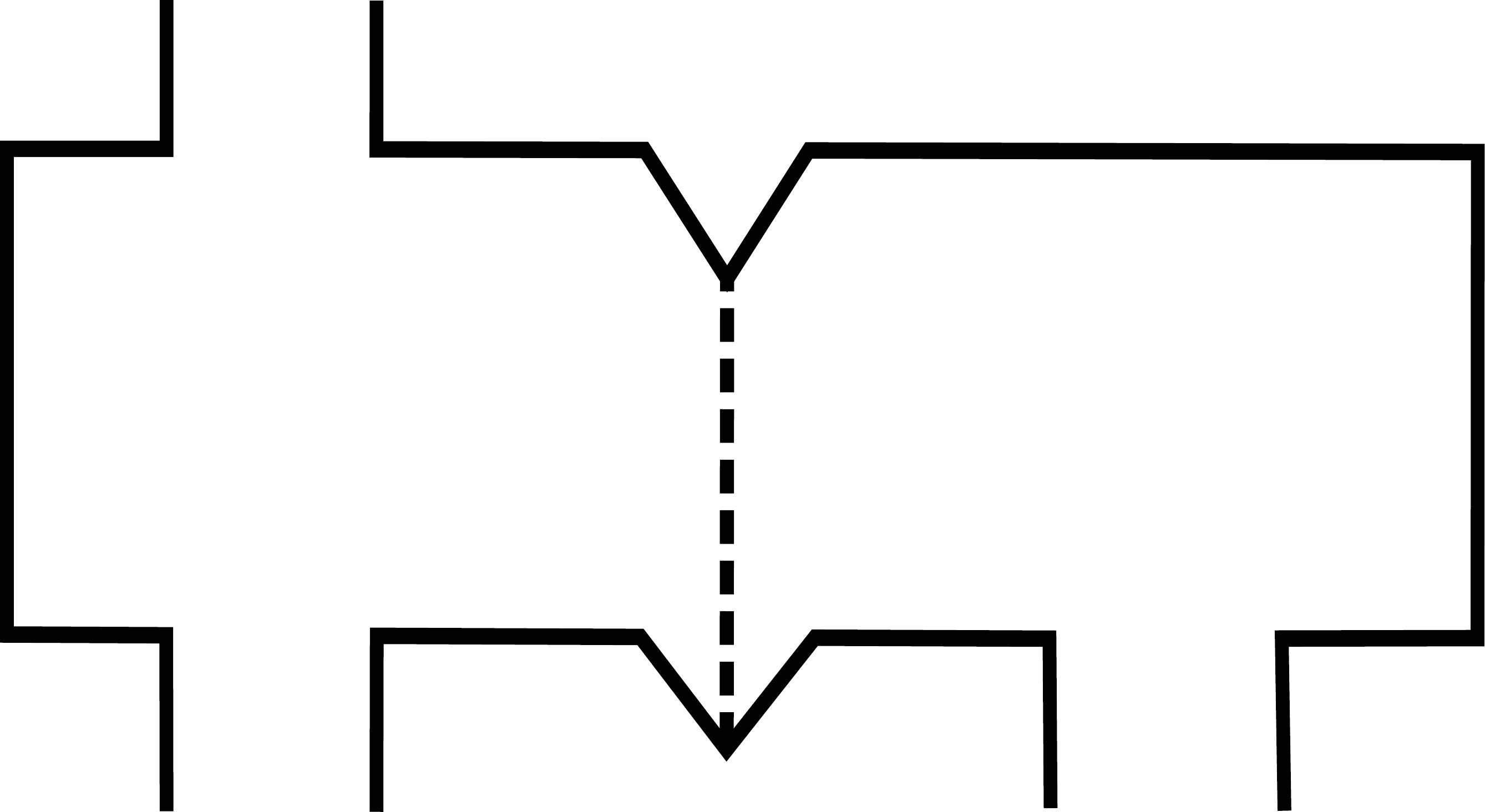}}
      & 0.02 & -- & 0.98 & -1 & W independent
            \\[0.7cm] \hline
\end{tabular}
      \caption{Transmissions $T_{10}$ and $T_{20}$ and polarizations
        $P_1$ and $P_2$ for different devices. The first column shows
        different sample geometries where the dashed line denotes the $p$-$n$ interface.
        (a) Transmissions for three different widths W for the same geometry as in Fig.~\ref{fig:1}, i.e. the $p$-$n$ interface meets armchair edges. The partitioning of the net transmission between $T_{10}$ and $T_{20}$ is a periodic
        function of $W$ with period $3a$ (or three hexagons across the
        width of the scattering region). (b) The $p$-$n$ interface meets edges of zigzag type. The two rows describe the zigzag/anti-zigzag configuration which lead to a different partitioning of the transmission. Note the changed position of the leads. (c) A triangular region is
        added to the lower edge (a zoom-in onto the tip is shown in
        the last column) to model different edge terminations. (d)--(e) Triangular
        regions added to the top and bottom edge. When the transmission to a particular lead is negligible, the polarization in this lead is not shown (long dash).}
      \label{tab:additional_cases}
      \end{center}
\end{table}

\begin{figure}
\includegraphics[width=0.79\columnwidth]{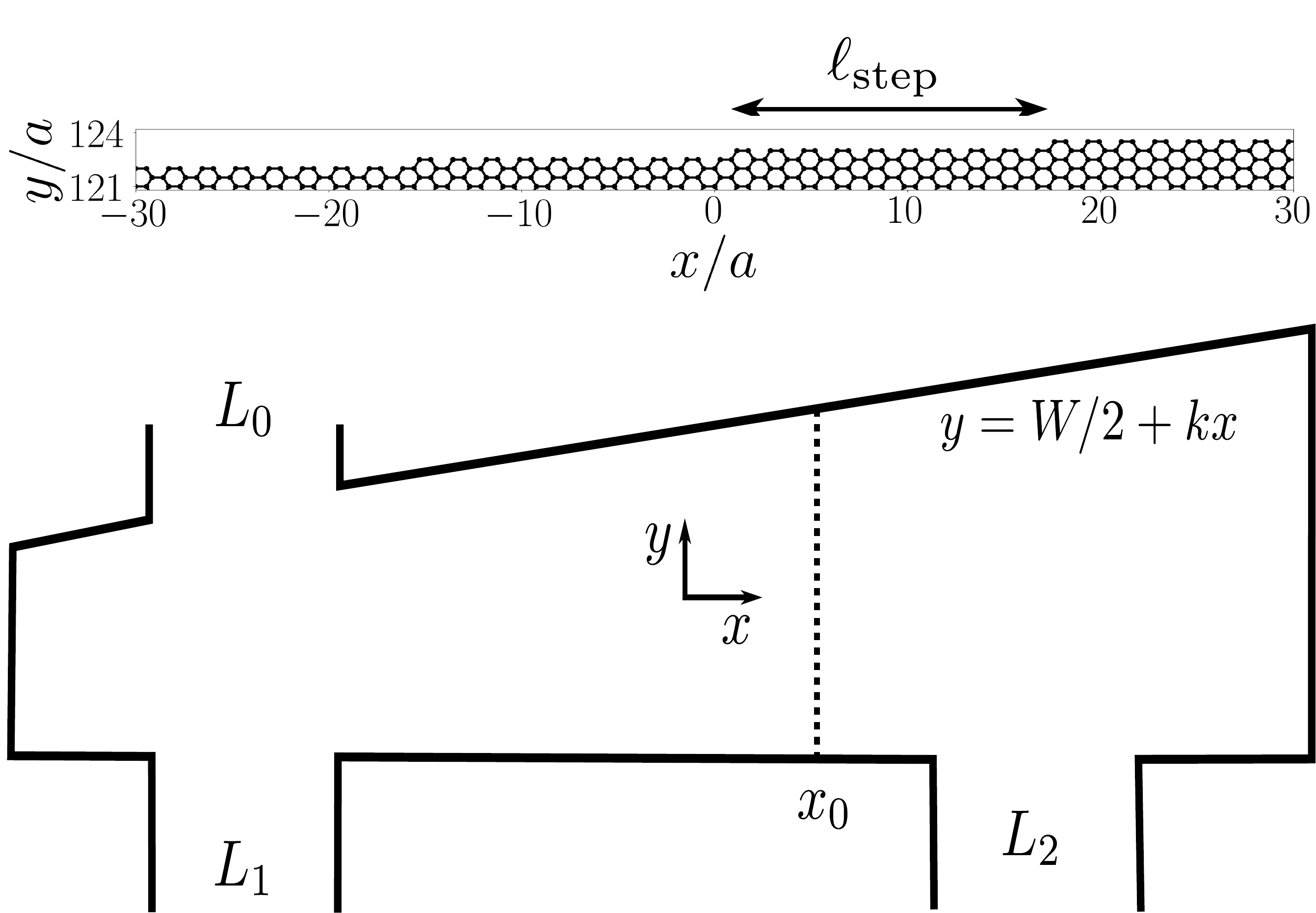}
\caption{Geometry of a device with a tilted staircase edge. Upper
  panel: zoom-in onto a part of the tilted upper edge displaying the
  staircase. The step length (length of the region of constant width
  $W$) is denoted by $\ell_\text{step}$. Lower panel: a schematic of
  the device. The slope of the tilted edge $k$ is related to
  $\ell_\text{step}$, e.g. $k=0.003$ corresponds to
  $\ell_\text{step}\approx 166a$ while $k=0.03$ corresponds to
  $\ell_\text{step}\approx 16a$. 
}
\label{fig:6}
\end{figure}

\begin{figure}
\centering
\subfloat[
]{\includegraphics[width=0.83\columnwidth]{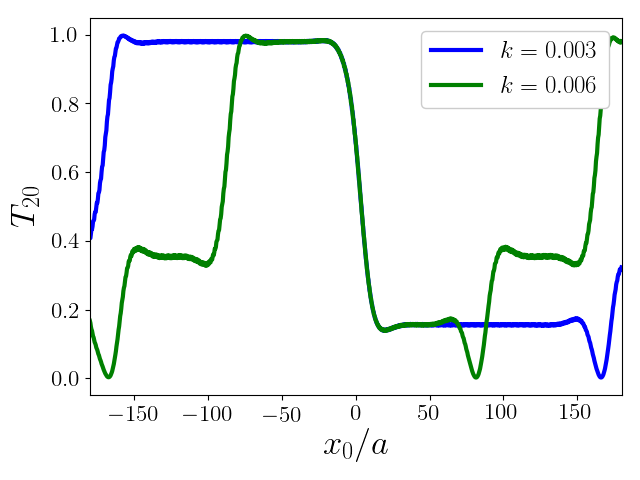}}\\
\renewcommand{\thesubfigure}{b}
\subfloat[
]{\includegraphics[width=0.83\columnwidth]{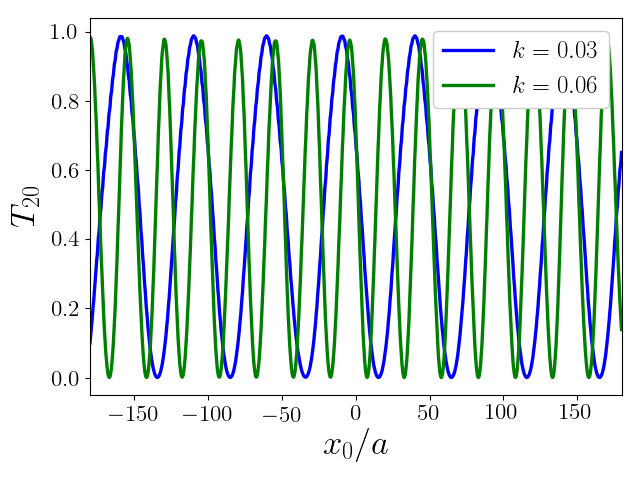}}
\caption{Transmission $T_{20}$ in a device with a tilted staircase
  edge as a function of the position of the $p$-$n$ interface $x_0$
  for different values of the slope $k$, see Fig.~\ref{fig:6}. (a)
  Plateau-like behavior of $T_{20}$ as expected for
  $\ell_\text{step}\gg 2\ell_B$. (b) Sine-like behavior of $T_{20}$
  for $\ell_\text{step}\lesssim 2\ell_B$. In this figure $L=780a$.}
\label{fig:7}
\end{figure}

\textit{Tilted staircase edge}. Now we consider the three-terminal
setup shown in Fig.~\ref{fig:6}. The upper edge has many steps on the
atomic scale, shown in the upper panel of Fig.~\ref{fig:6}. The size
of each of these steps is assumed to be constant and is denoted by
$\ell_{\text{step}}$.  The bottom-edge termination of the sample is of
armchair type.  We study the transmission $T_{20}$ as a function of
the position of the $p$-$n$ interface $x_0$. Note that
$T_{10}=1-T_{20}$, because here the parameters are such that only the
$n=0$ LL contributes to the electronic transport. If
$\ell_{\text{step}}\gg 2\ell_B$, the transmission $T_{20}$ shows a
plateau-like behavior (see Fig.~\ref{fig:7}(a)). The transmission
jumps to a different plateau as a new step is encountered while moving
$x_0$ from $-180 a$ to $180 a$ (the jump happens on length scales of
the order of $2\ell_B$). Since the upper and the lower edges are of
armchair type, we observe three plateau values corresponding to
different angles the between valley isospins at the two edges
$\Phi=\pi,\pm\pi/3$, in agreement with
Ref.~\onlinecite{Tworzydlo2007}. The width of these plateaus
corresponds to $\ell_{\text{step}}$. In the regime
$\ell_{\text{step}}\lesssim 2\ell_B$ there is a qualitative change
from the plateau-like to sine-like behavior of $T_{20}$, see
Fig.~\ref{fig:7}(b). Thus, when $\ell_{\text{step}}\lesssim 2\ell_B$,
the incoming current in $L_0$ can be partitioned into valley-polarized
currents in $L_1$ ($K$ valley) and $L_2$ (-$K$ valley) in any desired
ratio by tuning the location of the $p$-$n$ junction.  When
$\ell_{\text{step}} \lesssim 2\ell_B$, mixing of Landau orbits on
neighboring guiding centers gives rise to the conductance behavior
shown in Fig.~\ref{fig:7}(b). Our results are in agreement with the simulations 
in Ref.~\onlinecite{Handschin2017}.

In an experiment, one could measure the resulting valley polarization
by utilizing the valley Hall effect~\cite{Xiao2007}. This would
require breaking the inversion symmetry, which can be modeled by a
staggered sublattice potential of the form
$\pm\lambda_\nu\sum_i c^\dagger_i c_i$ in our system. Our results
remain valid even after adding such a term to the Hamiltonian in
Eq.~\eqref{eq:tb_hamiltonian} as long as $\lambda_\nu<V_0$.  This
condition ensures the presence of snake states in the system at
$E\approx 0$.

\section{Summary}
In summary, we have demonstrated that a graphene $p$-$n$ junction in a
uniform out-of-plane magnetic field can effectively function as a
switchable valley filter. The valley polarization of the carriers in
the outgoing leads is quite robust.  Changing the edge termination at
the $p$-$n$ interface can drastically modify the partitioning of the
current into the two outgoing leads, but the outgoing current in both
leads remains valley-polarized. We have also shown that in a device
where one of the edges has many steps on the atomic scale, the
partitioning of the current into two outgoing leads can be tuned by
choosing the \textit{location} of the $p$-$n$ junction. In such a
device it will be possible to partition a valley-unpolarized incoming
current into two streams of oppositely valley-polarized currents in
two outgoing leads in any desired ratio.

\section*{ACKNOWLEDGMENTS}
We would like to acknowledge fruitful discussions with P. Makk,
C. Handschin, and C. Sch\"{o}nenberger. This work
was financially supported by the Swiss National Science
Foundation (SNSF) and the NCCR Quantum Science and
Technology.  Work by EJM was supported by the Department of Energy,
Office of Basic Energy Sciences under grant DE FG02 84ER45118.

\end{document}